\documentclass[reprint,prb,aps,floatfix,superscriptaddress,nofootinbib,longbibliography]{revtex4-2}

\usepackage{amsmath,amssymb,graphicx,bm,color}
\usepackage[colorlinks=true,linkcolor=blue,citecolor=blue,urlcolor=blue]{hyperref}
\newcommand{\rs}{r_s}
\newcommand{\kF}{k_F}
\newcommand{\TF}{T_F}
\newcommand{\Nf}{N_F}
\newcommand{\Kxc}{K^{-}_{xc}}
\newcommand{\KxcPlus}{K^{+}_{xc}}
\newcommand{\chis}{\chi_s}
\newcommand{\chinaught}{\chi_0}

\newcommand{\thetared}{\theta}

\begin{document}

\title{Static spin exchange-correlation kernel of the uniform electron gas at finite temperature}

\author{Pengcheng Hou}
\thanks{These authors contributed equally.}
\affiliation{Hefei National Laboratory, University of Science and Technology of China, Hefei 230088, China}

\author{Zhiyi Li}
\thanks{These authors contributed equally.}
\affiliation{Department of Modern Physics, University of Science and Technology of China, Hefei, Anhui 230026, China}

\author{Youjin Deng}
\email{yjdeng@ustc.edu.cn}
\affiliation{Department of Modern Physics, University of Science and Technology of China, Hefei, Anhui 230026, China}
\affiliation{Hefei National Laboratory, University of Science and Technology of China, Hefei 230088, China}

\author{Kun Chen}
\email{chenkun@itp.ac.cn}
\affiliation{CAS Key Laboratory of Theoretical Physics, Institute of Theoretical Physics, Chinese Academy of Sciences, Beijing 100190, China}
\date{\today}

\begin{abstract}
    We determine the temperature and momentum dependence of the static spin
    exchange--correlation kernel of the unpolarized three-dimensional uniform
    electron gas using variational diagrammatic Monte Carlo.
    The calculations cover metallic densities with Wigner--Seitz radii
    $\rs=1$--$4$ and temperatures $T/\TF=1/16$--$8$, where $\TF$ is the
    Fermi temperature.
    At low temperature, the calculated kernel approaches established
    ground-state parametrizations.
    Heating suppresses its Fermi-scale minimum and reduces its magnitude,
    while the relative momentum dependence evolves nonmonotonically through
    the warm-dense crossover.
    At the two highest temperatures studied, the spin kernel is nearly
    constant over $q\leq2\kF$, where $\kF$ is the Fermi wave number.
    The charge kernel at $\rs=1$ retains pronounced momentum dependence
    under the same conditions, making a local approximation to the XC
    kernel more accurate in the spin channel over this interval.
    The directly sampled uniform response determines the XC free-energy
    curvature with respect to spin polarization, independently of a fitted
    polarization dependence.
    This curvature sets the interaction contribution to the free-energy
    cost of weak polarization and differs by several percent from the
    Groth--Dornheim--Bonitz parametrization at elevated temperatures.
    These results provide momentum-resolved reference data for finite-temperature
    spin-response theories and direct constraints on the spin dependence of
    thermal free-energy functionals.
\end{abstract}

\maketitle

\section{Introduction}
\label{sec:introduction}

Warm dense matter (WDM) encompasses compressed, partially degenerate
conditions encountered in inertial-confinement-fusion
plasmas~\cite{Hurricane2016}, giant-planet
interiors~\cite{SaumonGuillot2004,Schottler2018}, and laser-compressed
materials~\cite{Kraus2017}.
Quantitative modeling of this regime requires an accurate description
of electronic thermodynamics and response, accounting for thermal
excitation, Fermi statistics, and Coulomb interactions
~\cite{Dornheim2018,Bonitz2020}.
The three-dimensional uniform electron gas (UEG) provides a model for
studying these electronic effects in the absence of an ionic potential.
Its thermodynamic and response properties provide benchmarks for
many-body theories and for constructing and testing exchange-correlation
approximations~\cite{KSDT2014,Groth2017,Hou2022}.
The UEG is characterized by the Wigner--Seitz radius $\rs$ and the reduced
temperature $\thetared=T/\TF$, where $\TF$ is the Fermi temperature.

The charge and spin responses of the UEG describe changes in the total
electron density and in the relative spin populations, respectively.
Although the Coulomb interaction is spin independent, exchange and
correlations modify both responses
~\cite{GiulianiVignale2005,Dornheim2022spin}.
The resulting spin response cannot be inferred from the charge response
or unpolarized thermodynamics alone.
Describing weak spin polarization and its spatial variation therefore
requires a direct characterization of the spin channel.

The static spin susceptibility measures the response to a weak
spin-dependent field.
The static spin exchange--correlation (XC) kernel isolates the
interaction contribution to its inverse,
\begin{equation}
    \Kxc(q;\thetared)\equiv
    \chis^{-1}(q;\thetared)-\chinaught^{-1}(q;\thetared),
    \label{eq:Kxc-def}
\end{equation}
where $\chis$ and $\chinaught$ are the interacting and noninteracting
static spin susceptibilities, respectively, and $q$ is the wave number
of the perturbation. The density dependence is implicit.
A negative kernel corresponds to an enhanced spin susceptibility, with
$\chis/\chinaught=[1+\Kxc\chinaught]^{-1}$
~\cite{Stoner1938,Wolff1960}.
The uniform limit determines the XC contribution to the free-energy
cost of weak spin polarization at fixed total density
~\cite{GiulianiVignale2005}.
At finite momentum, the kernel describes how the interaction correction
changes with the spatial scale of the perturbation.
Its uniform value and finite-momentum shape thus characterize the
thermodynamic strength and spatial dependence of the same XC correction.
A local approximation replaces the momentum-dependent kernel by its uniform value.
Comparing the thermal evolution of the spin and charge kernels tests
whether this approximation describes the two channels equally well.

Quantum Monte Carlo (QMC) studies have established ground-state
spin-polarization energies~\cite{Spink2013} and finite-temperature thermodynamics
~\cite{Brown2013,Schoof2015,Malone2016,Dornheim2016TDL}.
These energetic data and theoretical limits have been used to construct
thermal XC free-energy parametrizations
~\cite{KSDT2014,Groth2017,Karasiev2018,Karasiev2019}.
Ground-state response data constrain spin and charge local-field-factor
models~\cite{KukkonenChen2021,KaplanKukkonen2023}.
At finite temperature, path-integral Monte Carlo (PIMC) calculations
have determined static charge responses and local-field corrections
over a range of densities and wave numbers
~\cite{Dornheim2017PBresponse,Groth2017response,Dornheim2019LFC}.
Hou \emph{et al.} calculated the finite-temperature charge XC kernel
using variational diagrammatic Monte Carlo (VDMC)~\cite{Hou2022}.
Response data provide tests of dielectric descriptions
~\cite{KukkonenOverhauser1979,Dornheim2022spin} and of kernels derived from
approximate density functionals~\cite{Moldabekov2023}.

For the spin channel, Chen and Haule reported low-temperature static
susceptibilities and local-field factors~\cite{ChenHaule2019}.
Li \emph{et al.} calculated the Matsubara-frequency-resolved kernel
at effectively zero temperature and small wave numbers, including the
ultranonlocal contribution associated with spin Coulomb
drag~\cite{Li2025}. At finite temperature, PIMC calculations determined
spin-resolved static responses at selected state
points~\cite{Dornheim2022spin}.
The quantum Singwi--Tosi--Land--Sj\"olander (qSTLS) approximation has been used
to study spin-resolved response over a broader parameter
range~\cite{AroraMoudgil2025}.
Complementary studies address spin-polarized momentum distributions and
the performance of density functionals
~\cite{Dornheim2021SpinPolarizedMomentum,Moldabekov2022SpinBenchmark}.
A systematic finite-temperature calculation is needed to follow the
spin kernel over Fermi-scale wave numbers and compare its thermal
evolution with the charge channel.

The thermodynamic information in the uniform response is directly
relevant to thermal spin-density functionals.
Within finite-temperature density-functional theory
(DFT)~\cite{Mermin1965,Stoitsov1988}, the local-spin-density approximation
(LSDA) uses the XC free energy of the UEG at the local density and spin polarization
~\cite{KSDT2014,Groth2017}.
Derivatives of this free energy determine the spin-dependent XC
potential and its response to changes in spin polarization
~\cite{RungeGross1984,GrossKohn1985,PribramJones2016}.
Near an unpolarized state, the second derivative with respect to
polarization is the XC spin curvature and determines the uniform spin kernel.
Describing weakly polarized states therefore requires an accurate curvature
as well as accurate free-energy values.
The need to test thermodynamic derivatives separately from energy values
is illustrated by the specific-heat anomalies discussed by Karasiev
\emph{et al.}~\cite{Karasiev2019}.

The Groth--Dornheim--Bonitz (GDB)~\cite{Groth2017} and
Karasiev--Sjostrom--Dufty--Trickey (KSDT)~\cite{KSDT2014}
parametrizations describe the UEG XC free energy using QMC data and
theoretical constraints.
Their analytical spin interpolations connect the unpolarized and fully
polarized free energies~\cite{PerrotDharmawardana2000} and determine
the curvature near zero polarization.
The corrected KSDT (corrKSDT) parametrization updates the unpolarized
free energy~\cite{Karasiev2018}; its agreement with GDB for this quantity
does not establish agreement of their spin curvatures.
GDB includes finite-temperature QMC data at partial polarization, but
the zero-polarization curvature remains a prediction of its fitted form.
Direct uniform-response data provide an independent test of this prediction.

In this work, we calculate the static spin XC kernel through fifth order
in VDMC over metallic densities and a temperature range extending from
quantum degeneracy to nondegenerate conditions.
The low-temperature results approach established ground-state
parametrizations and earlier VDMC data.
Heating suppresses the Fermi-scale minimum and reduces the kernel
magnitude, but the relative momentum dependence evolves nonmonotonically
through the warm-dense crossover.
At $\thetared=4$ and $8$, the spin kernel is nearly constant for wave
numbers up to twice the Fermi wave number at all studied densities.
Matched charge-channel data at $\rs=1$ retain appreciable momentum
dependence under the same conditions.
This contrast remains after removing the bare Coulomb contribution from
the charge kernel and therefore concerns the XC corrections themselves.
It identifies a regime in which a local representation describes the
spin XC correction more accurately than its charge-channel counterpart.

The directly calculated uniform response determines the XC
spin curvature without assuming a fitted polarization dependence.
It provides an independent benchmark for thermal free-energy
parametrizations, resolving differences of several percent from GDB
at elevated temperatures.
GDB and corrKSDT, the latter evaluated with the original KSDT spin
interpolation, predict substantially different spin curvatures despite
their close agreement for the unpolarized free energy.
An analysis of the endpoint functions and spin interpolations traces
this high-temperature discrepancy primarily to the polarized--unpolarized
free-energy differences.
Comparison with first-order exchange shows that the nearly local
high-temperature kernel retains an appreciable correlation contribution.
The momentum-resolved and uniform results thus provide complementary
constraints on static spin-response theories and the polarization
dependence of thermal free energies.

Section~\ref{sec:methods} introduces the model and computational method.
Section~\ref{sec:results} presents the momentum-resolved kernel, the
spin--charge comparison, and the uniform spin-curvature results.
Section~\ref{sec:conclusions} summarizes the findings and their implications.
The Appendixes present convergence tests, the density dependence of the
normalized kernel, and the exchange reference and thermal free-energy
curvature analysis.

\section{Model and Methods}
\label{sec:methods}

We consider the three-dimensional unpolarized UEG in a homogeneous
neutralizing positive background.  In Rydberg atomic units,
$\hbar=2m_e=e^2/2=k_B=1$, the grand-canonical Hamiltonian is
\begin{equation}
    \begin{aligned}
        \hat{\mathcal H}={} & \sum_{\mathbf k,\sigma}(k^2-\mu)
        c^\dagger_{\mathbf k\sigma}c_{\mathbf k\sigma}                          \\
                            & +\frac{1}{2V}\sum_{\mathbf q\ne0}\frac{8\pi}{q^2}
        \sum_{\mathbf k,\mathbf k',\sigma,\sigma'}
        c^\dagger_{\mathbf k+\mathbf q,\sigma}
        c^\dagger_{\mathbf k'-\mathbf q,\sigma'}
        c_{\mathbf k'\sigma'}c_{\mathbf k\sigma}.
    \end{aligned}
    \label{eq:ueg-hamiltonian}
\end{equation}
Here $c^\dagger_{\mathbf k\sigma}$ and $c_{\mathbf k\sigma}$ create and
annihilate an electron with momentum $\mathbf k$ and spin $\sigma$,
$V$ is the volume, and $\mu$ fixes the mean density.  The background
removes the $\mathbf q=0$ Coulomb term.  Lengths are measured in Bohr
radii $a_0$, and $\rs$ is the radius of a sphere containing one electron,
in units of $a_0$. The number density is $n=3/(4\pi\rs^3)$, with Fermi wave number
$\kF=(9\pi/4)^{1/3}/\rs$ and $\TF=\kF^2$ in these units.

The relative spin polarization is $\zeta=(n_\uparrow-n_\downarrow)/n$,
where $n_\uparrow$ and $n_\downarrow$ are the spin-resolved densities.
We define the linear spin response by
$\delta(n_\uparrow-n_\downarrow)_q=\chis(q)h_q$,
where the static field $h(\mathbf r)$ shifts the up- and down-spin
one-particle energies by $-h(\mathbf r)$ and $+h(\mathbf r)$, respectively.
With this convention,
$\chinaught(0;T=0)=\Nf=\kF/(2\pi^2)$, the zero-temperature ideal-gas
density of states per unit volume, including both spin species.
We express the kernel in dimensionless form as $\Kxc\Nf$.
Here $\Nf$ is only a fixed zero-temperature response scale; the
$\chinaught(q;T)$ entering Eq.~\eqref{eq:Kxc-def} is the finite-temperature
Lindhard susceptibility. The unscaled kernel has units $\mathrm{Ry}\,a_0^3$.

We calculate the spin susceptibility at zero Matsubara frequency in the
thermodynamic limit using VDMC~\cite{ChenHaule2019}.
The diagrammatic expansion is organized around a propagator containing
the Fock self-energy and the screened interaction $8\pi/(q^2+\lambda)$.
Here $\lambda$ is an auxiliary screening parameter with units $a_0^{-2}$.
Interaction counterterms restore the bare Coulomb interaction in the
full expansion, while chemical-potential counterterms maintain the
target density order by order.

We retain the series through fifth order and combine the sampled
corrections with the analytically evaluated noninteracting terms to
obtain the full susceptibility.
The spin XC kernel is then calculated using Eq.~\eqref{eq:Kxc-def}
with the finite-temperature Lindhard response.
The uniform response is sampled directly.
At each momentum, we select the available $\lambda$ that minimizes the
absolute difference between the fourth- and fifth-order kernels.
The error bars combine the propagated Monte Carlo uncertainty and this
last-order difference in quadrature, with the latter serving as a
truncation-error estimate.
Convergence tests and the uncertainty analysis are presented in
Appendix~\ref{app:vdmc}.

\section{Results}
\label{sec:results}

Temperature changes both the magnitude and the momentum dependence of
the static spin XC kernel.
We first characterize this evolution, then compare the relative momentum
dependence of the spin and charge channels and assess the uniform spin
curvature against thermal free-energy parametrizations.
The calculations cover $\rs=1,2,3,4$ at eight reduced temperatures,
$\thetared=1/16,1/8,1/4,1/2,1,2,4,8$.
The wave-number grid contains 13 points, $q/\kF=0,0.2,\ldots,2.4$,
spanning the uniform limit and the Fermi-scale region.

\subsection{Momentum-resolved kernel}
\label{sec:momentum}

\begin{figure*}[!t]
    \centering
    \includegraphics[width=\textwidth]{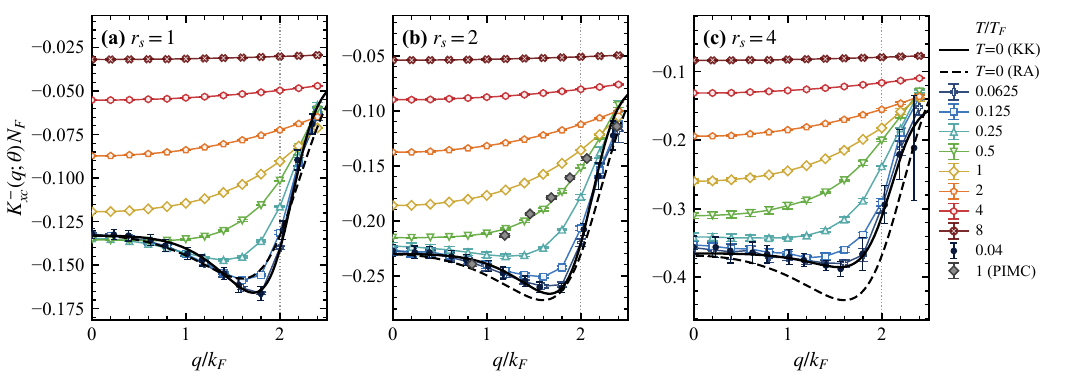}
    \caption{\label{fig:kxc_vs_q}
        Static spin XC kernel $\Kxc(q;\thetared)\Nf$ versus $q/\kF$
        at $\rs=1$ (a), $2$ (b), and $4$ (c).
        Open colored markers show the present VDMC results for
        $\thetared=1/16$--$8$; filled navy circles are the
        $\thetared=0.04$ data of Ref.~\cite{ChenHaule2019}.
        Black solid and dashed curves are the zero-temperature
        Kaplan--Kukkonen (KK)~\cite{KaplanKukkonen2023} and
        Richardson--Ashcroft (RA)~\cite{RichardsonAshcroft1994} parametrizations.
        Gray diamonds in panel (b) are reconstructed from the PIMC data
        for $N=14$ electrons at $\rs=2,\thetared=1$ in
        Refs.~\cite{Dornheim2022spin,Dornheim2022rodare}.
        Error bars for the present VDMC results denote the combined
        uncertainties defined in Sec.~\ref{sec:methods};
        those of Ref.~\cite{ChenHaule2019} are statistical uncertainties.
        PIMC conversion and error propagation are described in
        Appendix~\ref{app:vdmc}.
        Lines connecting markers guide the eye.}
\end{figure*}

Figure~\ref{fig:kxc_vs_q} shows the static spin XC kernel at three
representative densities, $\rs=1,2,4$.
The kernel is negative throughout the displayed range, corresponding
to an enhancement of the spin susceptibility.
At the lowest temperatures, it is nearly constant at small momentum
and has a broad negative minimum around $q\simeq1.5$--$2\kF$.
The lowest-temperature results closely follow the zero-temperature
Kaplan--Kukkonen parametrization and the earlier VDMC data at
$\thetared=0.04$~\cite{ChenHaule2019} over $q\leq2\kF$.
The Richardson--Ashcroft parametrization reproduces the broad minimum
but predicts a deeper minimum at $\rs=4$.
The low-momentum kernel changes little over the lowest temperatures
studied, before its magnitude decreases on further heating.
Our calculations are performed directly in the thermodynamic limit,
with representative order-convergence and screening-parameter tests
presented in Appendix~\ref{app:vdmc}.

Heating suppresses the broad Fermi-scale minimum, which remains visible
at $\thetared=0.25$ but is progressively smoothed for
$\thetared\gtrsim0.5$.
Further heating reduces the kernel magnitude across the plotted momenta.
The larger magnitude of $\Kxc\Nf$ at larger $\rs$ reflects a stronger
XC correction relative to the zero-temperature ideal-gas response scale.
This density dependence is visible both in the low-momentum values and
in the depth of the minimum.

At $\rs=2$ and $\thetared=1$, the kernel reconstructed from the published
spin-resolved PIMC data~\cite{Dornheim2022spin} is negative and decreases
in magnitude with momentum, as in our VDMC results, but its magnitude
is systematically larger [Fig.~\ref{fig:kxc_vs_q}(b)].
At the lowest available PIMC momentum, $q/\kF\simeq0.84$, the magnitude
remains comparable to the zero-temperature parametrizations.
In contrast, our VDMC temperature series shows substantial thermal
weakening at this momentum by $\thetared=1$.
The PIMC magnitude exceeds the VDMC value by about $33\%$ at this
momentum, with the difference decreasing to $7\%$ near $q/\kF=2.38$.

Across the studied densities, the spin XC kernel evolves from a
low-temperature profile with a pronounced Fermi-scale minimum to a
weaker, smoother form at elevated temperatures.

\subsection{Spin--charge contrast in momentum dependence}
\label{sec:flattening}

\begin{figure}[!htbp]
    \centering
    \includegraphics[width=0.85\columnwidth]{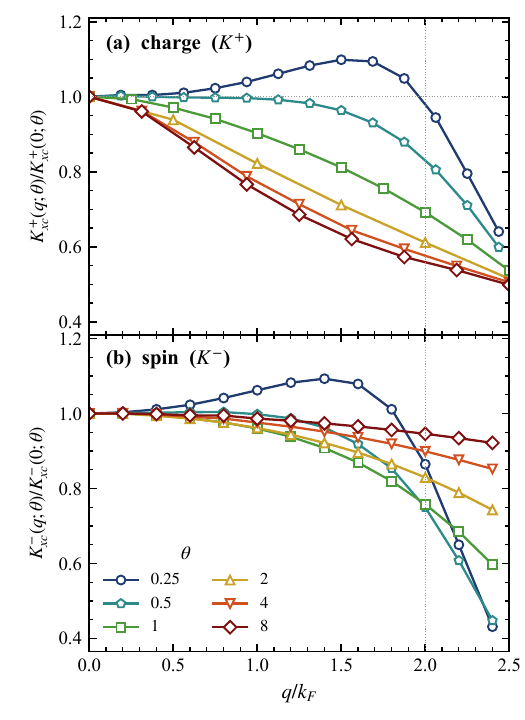}
    \caption{\label{fig:locality}
        Normalized static XC kernels at $\rs=1$: (a) charge,
        $\KxcPlus(q;\thetared)/\KxcPlus(0;\thetared)$, and (b) spin,
        $\Kxc(q;\thetared)/\Kxc(0;\thetared)$, for
        $\thetared=0.25,0.5,1,2,4,8$.
        Charge-channel data are from our earlier VDMC study~\cite{Hou2022},
        supplemented by an additional present calculation at $\thetared=0.5$.
        Spin-channel data are from the present work.
        Horizontal and vertical dotted lines mark unity and $q=2\kF$,
        respectively.
        Lines connect normalized central values; the kernel uncertainties are
        discussed in Appendix~\ref{app:vdmc}.}
\end{figure}

We separate changes in the kernel shape from its overall thermal
weakening by considering
$\Kxc(q;\thetared)/\Kxc(0;\thetared)$.
Unity corresponds to a momentum-independent kernel fixed to the
calculated uniform value.
Figure~\ref{fig:locality} compares the two channels at $\rs=1$ and
$\thetared=0.25,0.5,1,2,4,8$, using charge data from our earlier VDMC
study~\cite{Hou2022} and additional results at $\thetared=0.5$.
Each channel is normalized by its own uniform kernel.
The charge XC kernel $\KxcPlus$ excludes the bare Coulomb contribution,
so the comparison concerns the momentum dependence of the XC
corrections themselves.

The two channels show a pronounced contrast at high temperature.
At $q=2\kF$, the spin ratio is about $0.90$ and $0.95$ for
$\thetared=4$ and $8$, respectively, whereas the corresponding charge
ratios are about $0.58$ and $0.56$.
This contrast supports a more accurate local representation of the
spin XC correction over the studied momentum interval, while the
charge XC correction retains substantial spatial dependence.
The weak spin-kernel momentum dependence extends across the studied
densities, as shown in Appendix~\ref{app:density-scan}.
Across $\rs=1$--$4$, the calculated spin profiles vary by at most
about $12\%$ at $\thetared=4$ and $6\%$ at $\thetared=8$
relative to their uniform values over $q\leq2\kF$.

At $\rs=1$, the normalized spin kernel varies nonmonotonically with temperature.
At $q=2\kF$, the ratio decreases from about $0.86$ at $\thetared=0.25$
to $0.75$ near $\thetared=0.5$--$1$, before increasing at higher temperatures.
Initially, the finite-momentum kernel weakens while the uniform
magnitude changes little.
At higher temperatures, the uniform magnitude decreases more rapidly
in relative terms, bringing the ratio closer to unity.
The overall thermal weakening of the kernel therefore accompanies an
initial increase, followed by a decrease, in its relative momentum
variation at $q=2\kF$.

The thermal momentum scale $\sqrt{T}$ grows relative to $\kF$ as
the temperature increases, so a fixed interval in $q/\kF$ probes
progressively smaller values of $q/\sqrt{T}$ in both channels.
In a strictly classical system with spin-independent interactions,
the interaction contribution to the free energy is independent of spin polarization.
The spin response then obeys the Curie law $\chis=\chinaught=n/T$,
with a vanishing spin XC kernel
~\cite{GiulianiVignale2005,PerrotDharmawardana2000}.
At finite temperature, exchange and quantum correlations leave a
nonzero spin XC correction whose spatial dependence must be determined
from the response.
The channel comparison shows that its relative momentum variation is
already small over Fermi-scale wave numbers at $\thetared=4,8$,
while that of the charge XC correction remains substantial.
A local spin kernel in this regime still produces a momentum-dependent
susceptibility through the Lindhard response.
The accuracy of a local approximation therefore depends on both the
residual momentum dependence and the uniform value assigned to the kernel.

\subsection{Uniform spin curvature and thermal free-energy parametrizations}
\label{sec:stiffness}

For the XC free energy per electron $f_{xc}(\rs,\thetared,\zeta)$,
the leading change induced by a weak uniform polarization is
$\tfrac12\alpha_{xc}\zeta^2$ at fixed density and temperature.
The coefficient $\alpha_{xc}=\partial^2 f_{xc}/\partial\zeta^2|_{\zeta=0}$
is the XC spin curvature.
It determines the interaction contribution to the uniform inverse
susceptibility through $\Kxc(0;\thetared)=\alpha_{xc}/n$
~\cite{GiulianiVignale2005,PerdewWang1992}.
We obtain this coefficient directly from the uniform response,
independently of a fitted polarization dependence.
A negative XC curvature lowers the free-energy cost of polarization
and enhances the susceptibility.
In thermal LSDA, the polarization curvature of the input free energy,
divided by $n$, supplies a momentum-independent spin kernel
~\cite{KSDT2014,Groth2017}.
The comparison at $q=0$ therefore tests the parametrized curvature
independently of the local spatial approximation.

\begin{figure}[!t]
    \centering
    \includegraphics[width=\columnwidth]{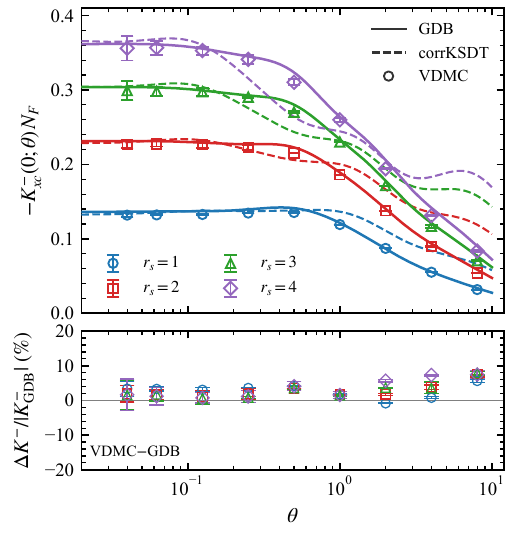}
    \caption{\label{fig:chi_ratio}
    Uniform spin XC kernel and comparison with thermal free-energy parametrizations.
    Top: $-\Kxc(0;\thetared)\Nf$ at $\rs=1,2,3,4$.
    Solid and dashed curves are obtained from GDB and corrKSDT,
    respectively; symbols denote VDMC, with the $\thetared=0.04$ points
    from Ref.~\cite{ChenHaule2019}.
    Bottom: $(K_{xc}^--K_{xc,\mathrm{GDB}}^-)/|K_{xc,\mathrm{GDB}}^-|$,
    expressed as a percentage.
    Error bars for the present results are defined in Sec.~\ref{sec:methods};
    those of Ref.~\cite{ChenHaule2019} are statistical uncertainties.
    Uncertainties of the fitted parametrizations are not included.}
\end{figure}

We compare the uniform kernel with GDB and corrKSDT in
Fig.~\ref{fig:chi_ratio}.
For corrKSDT, we use the corrected unpolarized free energy together
with the original KSDT fully polarized endpoint and spin interpolation
~\cite{Karasiev2018,KSDT2014}; the endpoint functions and temperature
conventions are specified in Appendix~\ref{app:lsda}.
The upper panel shows the thermal evolution at all four densities.
At $\rs=1$, the magnitude exhibits a shallow maximum near
$\thetared=0.5$. At the larger $\rs$ values it generally decreases
with temperature, and all four densities show a rapid decrease beyond
$\thetared\simeq0.5$--$1$.
The dimensionless magnitude increases with $\rs$ at fixed reduced temperature.
At low temperature, the present values approach the earlier VDMC results,
consistent with the low-momentum behavior in Fig.~\ref{fig:kxc_vs_q}.

The lower panel of Fig.~\ref{fig:chi_ratio} resolves the differences
between the VDMC and GDB uniform kernels.
The VDMC--GDB differences have magnitudes of about $1$--$7\%$ at
$\thetared=2$--$4$.
They are positive at these state points except for a small negative
difference at $\rs=1,\thetared=2$.
At $\thetared=8$, the VDMC curvature magnitudes lie about $6$--$8\%$
below the GDB predictions across the four densities.
At the same temperature, corrKSDT predicts curvature magnitudes
approximately twice those of GDB.

At high temperature, the ideal-gas contribution dominates the inverse
susceptibility, while the XC kernel isolates the spin-dependent
interaction correction.
For example, at $\rs=1,\thetared=8$, the VDMC and GDB kernels differ
by about $6\%$, while the corresponding full susceptibilities differ
by less than $0.02\%$.
The kernel comparison directly tests this interaction contribution,
whose relative accuracy is difficult to assess from the full response alone.

The exchange and correlation contributions to this curvature can be
separated using the first-order exchange reference in Appendix~\ref{app:lsda}.
The correlation contribution, $K_c^-(0)=\Kxc(0)-K_x^-(0)$, is positive
throughout the studied grid and offsets $21$--$48\%$ of the exchange
magnitude at $\thetared=8$ [Table~\ref{tab:exchange-reference}].
The nearly local high-temperature kernel thus retains an appreciable
correlation contribution to its uniform value.

To understand the difference between GDB and corrKSDT, we separate
the roles of their endpoint free energies and spin interpolation.
The Perrot--Dharma--Wardana (PDW) interpolation used for both
connects the unpolarized and fully polarized states
~\cite{PerrotDharmawardana2000}.
Its curvature at zero polarization, multiplied by the endpoint
free-energy difference, gives the XC spin curvature.
The two unpolarized free energies agree closely~\cite{Karasiev2019},
but the polarized--unpolarized difference is much smaller than either
endpoint free energy and is more sensitive to differences between the fits.
At $\thetared=8$, the interpolation curvature factors are nearly equal
in the two parametrizations [Fig.~\ref{fig:pdw-diagnostic}].
Their different spin-curvature predictions therefore originate primarily
in the polarized--unpolarized free-energy differences.

Uniform-response data fix the curvature at zero polarization, whereas
QMC free energies constrain the polarization dependence over a finite interval.
Together, they constrain the endpoint functions and spin interpolation
of thermal free-energy parametrizations.

\section{Conclusions}
\label{sec:conclusions}

We have determined the static spin XC kernel of the unpolarized
three-dimensional UEG at metallic densities from the quantum-degenerate
to the nondegenerate regime.
The kernel weakens and loses its Fermi-scale minimum with heating,
while its normalized momentum dependence evolves nonmonotonically.
At $\thetared=4$ and $8$, the spin kernel is nearly constant over
$q\leq2\kF$ at all four densities, whereas the matched charge kernel
at $\rs=1$ retains appreciable momentum dependence.
This contrast supports an approximately local spin XC correction on
these scales, while the charge correction retains substantial spatial dependence.

The uniform response independently determines the XC free-energy curvature
with respect to spin polarization, providing a direct benchmark for
thermal free-energy parametrizations.
At elevated temperatures, the response-derived curvature differs by
several percent from GDB.
The comparison with corrKSDT shows that closely agreeing unpolarized
free energies can yield
substantially different spin curvatures.
The high-temperature difference between the parametrizations originates
primarily in their polarized--unpolarized free-energy differences.
The exchange reference shows that the nearly local high-temperature
kernel still requires an appreciable correlation contribution to its
uniform value.

The calculated kernels provide reference data for finite-temperature
theories of the static spin response over the sampled density,
temperature, and wave-number range.
The momentum-resolved data assess the spatial approximation in thermal LSDA,
while the uniform response tests the spin curvature supplied by its
free-energy parametrization.
Combining uniform-response constraints with free energies at partial
polarization can guide improvements to the endpoint functions and spin
interpolation of thermal parametrizations.

Comparison with the available finite-temperature PIMC reconstruction
reveals a systematic difference in the kernel magnitude while preserving
the same sign and overall momentum trend.
The thermodynamic-limit VDMC data, with a directly sampled uniform
limit and a systematic temperature series, provide an independent
reference for further comparisons of the spin-resolved responses.
Future calculations at larger wave numbers can test the extent of the
observed flattening, while controlled high-temperature expansions can
identify the terms governing the residual momentum dependence.

\begin{acknowledgments}
    P.H., Z.L., and Y.D. were supported by the National Natural Science Foundation
    of China (Grant No. 12275263) and the Quantum Science and Technology--National
    Science and Technology Major Project (Grant No. 2021ZD0301900).  K.C. was
    supported by the National Key Research and Development Program of China,
    Grant No. 2024YFA1408604, the National Natural Science Foundation of China
    under Grants No. 12474245 and No. 12447103, and the GHfund A(202407010637).
\end{acknowledgments}

\paragraph*{Data availability.}
The spin XC-kernel tables with combined uncertainties, representative
convergence data, charge-channel references, and uniform-curvature,
first-order exchange, and fixed-endpoint diagnostic tables are available
in the public repository of Ref.~\cite{UEGSpinData}.
The earlier $\thetared=0.04$ spin-response data are reported in
Ref.~\cite{ChenHaule2019}; the PIMC data used for comparison are available
from Refs.~\cite{Dornheim2022spin,Dornheim2022rodare}.

\appendix

\section{Convergence and uncertainty estimates}
\label{app:vdmc}

\begin{figure*}[!t]
    \centering
    \includegraphics[width=0.95\textwidth]{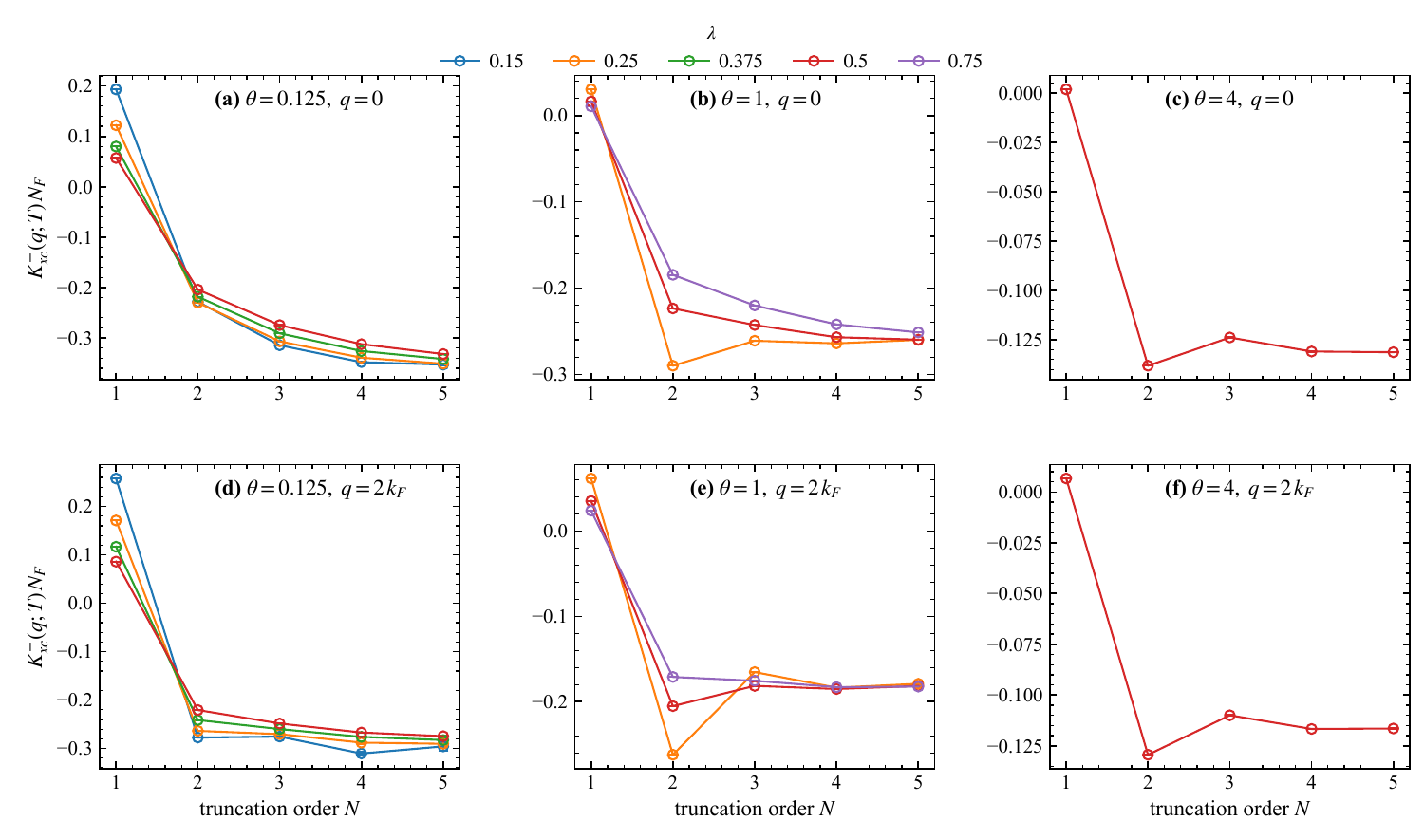}
    \caption{\label{fig:app-convergence}
    Order dependence of $\Kxc(q;\thetared)\Nf$ at $\rs=4$,
    for $q=0$ (top) and $q=2\kF$ (bottom).
    The columns correspond to $\thetared=0.125$, $1$, and $4$.
    Curves show kernels obtained from successive response partial sums at
    representative fixed screening parameters $\lambda$, in $a_0^{-2}$;
    at $\thetared=4$ the calculation
    uses $\lambda=0.5$.
    Error bars show propagated statistical uncertainties.}
\end{figure*}

Figure~\ref{fig:app-convergence} illustrates the order convergence of
the spin XC kernel at $\rs=4$, the largest coupling parameter studied,
for both the uniform response and $q=2\kF$.
Each curve is obtained by inverting successive partial sums of the full
response at a fixed screening parameter.
At $\thetared=0.125$, the convergence rate depends on $\lambda$, and
some residual order dependence remains at fifth order.
At $\thetared=1$, the higher-order estimates for different $\lambda$
approach similar values, while the fourth- and fifth-order results
closely agree at $\thetared=4$.

The reported kernels use the screening parameter with the smallest
fourth-to-fifth-order change at each momentum, as specified in
Sec.~\ref{sec:methods}.
The uncertainty of each final kernel combines the propagated statistical
uncertainty with this estimate of the truncation error in quadrature.
The normalized profiles are formed from the same pointwise-selected
kernel values.

For $\rs=2,\thetared=1$, the state compared with PIMC in
Fig.~\ref{fig:kxc_vs_q}(b), Table~\ref{tab:pimc-state-convergence}
lists the third- through fifth-order kernels at four representative
wave numbers, showing the stability of the higher-order results.
At $\thetared=4$ and $8$, only one screening parameter was calculated
for each density, with the same $\lambda$ used throughout its momentum grid.
Across these eight high-temperature states and $q\leq2\kF$, the
fourth-to-fifth-order change is below $1\%$ of the kernel magnitude
and smaller than the preceding order increment.
The normalized profiles are likewise stable between the two highest
orders, supporting the weak momentum dependence reported in
Sec.~\ref{sec:flattening}.

\begin{table}[!htbp]
    \caption{\label{tab:pimc-state-convergence}
    Order dependence of $\Kxc\Nf$ at $\rs=2,\thetared=1$.
    Each row uses the same $\lambda$, in $a_0^{-2}$, for all three
    orders, selected by the criterion in Sec.~\ref{sec:methods}.
    Parentheses on the fifth-order values denote the combined
    uncertainty in the last digits.}
    \begin{ruledtabular}
        \begin{tabular}{ccrrr}
            $q/\kF$ & $\lambda$ & Third order & Fourth order & Fifth order    \\
            \hline
            $0.0$   & $0.75$    & $-0.18366$  & $-0.18616$   & $-0.18605(20)$ \\
            $0.8$   & $0.75$    & $-0.17894$  & $-0.18070$   & $-0.18039(32)$ \\
            $1.6$   & $0.75$    & $-0.15835$  & $-0.15835$   & $-0.15773(63)$ \\
            $2.4$   & $0.75$    & $-0.10643$  & $-0.10573$   & $-0.10540(38)$ \\
        \end{tabular}
    \end{ruledtabular}
\end{table}

For the PIMC comparison in Fig.~\ref{fig:kxc_vs_q}(b), we convert the
published spin-resolved local-field factors~\cite{Dornheim2022spin,Dornheim2022rodare}
using $\Kxc=-4\pi(G_{\uparrow\uparrow}-G_{\uparrow\downarrow})/q^2$
in the units of Sec.~\ref{sec:methods}.
We propagate the reported channel uncertainties without a cross-channel
covariance and apply no additional finite-size correction.
The relative differences in Sec.~\ref{sec:momentum} are calculated by
linearly interpolating the VDMC kernel onto the PIMC wave numbers
within the common interval.

\section{Density dependence of the normalized spin kernel}
\label{app:density-scan}

\begin{figure*}[!t]
    \centering
    \includegraphics[width=0.85\textwidth]{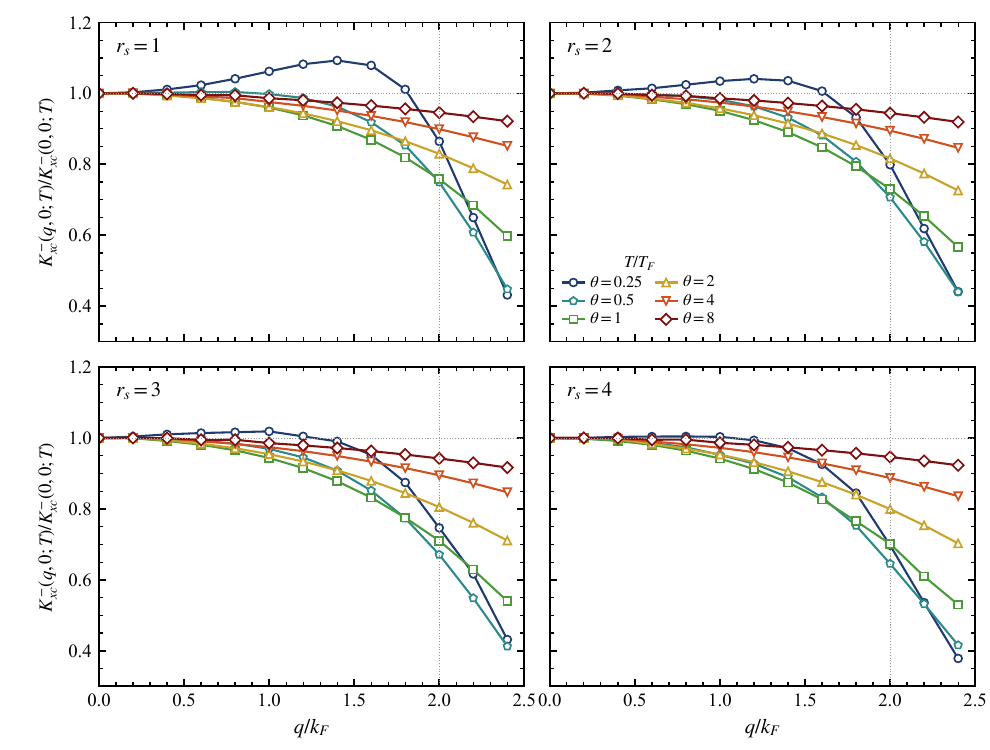}
    \caption{\label{fig:app-locality-rs}
        Normalized static spin XC kernel
        $\Kxc(q;\thetared)/\Kxc(0;\thetared)$ at $\rs=1,2,3,4$
        and $\thetared=0.25,0.5,1,2,4,8$.
        Curves show normalized central values; dotted lines mark unity and $q=2\kF$.
        The four panels use the same vertical scale.}
\end{figure*}

Figure~\ref{fig:app-locality-rs} extends the spin-kernel comparison in
Sec.~\ref{sec:flattening} to all four studied densities.
At $\thetared=0.25$, the modest maximum at intermediate wave number
becomes less pronounced as $\rs$ increases.
At all four densities the kernel remains close
to its uniform value at small momentum, while appreciable relative
variation develops toward $q\simeq2\kF$ at intermediate temperatures.

At $\thetared=4$ and $8$, all four normalized profiles are much flatter
over $q\leq2\kF$ and depend weakly on density.
The sampled profiles vary by at most about $12\%$ and $6\%$,
respectively, relative to their uniform values on this interval.
The weak density dependence of the normalized shape contrasts with
the distinct density dependence of the uniform amplitude
    [Fig.~\ref{fig:chi_ratio}].
The momentum variation increases beyond $2\kF$ at both temperatures.

\section{Uniform spin curvature: exchange reference and parametrizations}
\label{app:lsda}

\subsection{Exchange reference}

The exchange reference used in Sec.~\ref{sec:stiffness} is the
first-order contribution to the free energy at fixed density,
evaluated with ideal-gas occupations. In the units of Sec.~\ref{sec:methods},
the free energy per electron is
\begin{equation}
    f_x=-\frac{1}{2n}\sum_\sigma
    \int\!\frac{d^3k\,d^3p}{(2\pi)^6}
    \frac{8\pi\,f_\sigma(k)f_\sigma(p)}{|\mathbf k-\mathbf p|^2},
    \label{eq:exchange-reference}
\end{equation}
where $f_\sigma(k)=[1+\exp((k^2-\mu_{0\sigma})/T)]^{-1}$ and
$\mu_{0\sigma}$ fixes the ideal-gas density $n_\sigma$.
We obtain $K_x^-(0)=n^{-1}\partial_\zeta^2 f_x|_{\zeta=0}$,
holding the total density and absolute temperature fixed, consistent
with finite-temperature exchange spin scaling~\cite{KSDT2014}.

Subtracting this exchange reference from the VDMC uniform kernel gives
the positive correlation contribution $K_c^-(0)=\Kxc(0)-K_x^-(0)$.
On the sampled temperature grid, both the exchange magnitude and the
correlation contribution are largest at $\thetared=0.5$ for all four densities.
Their increases from the lowest temperature to $\thetared=0.5$ nearly
cancel at $\rs=1$, leaving the shallow maximum in the full kernel magnitude.
At $\rs=2$--$4$, the correlation increase exceeds the exchange increase,
so the full kernel instead decreases in magnitude over this interval.

At fixed $\thetared$, the exchange reference $K_x^-(0)\Nf$ scales
linearly with $\rs$.
An independent real-space evaluation provides a numerical cross-check
of the exchange integral.
The zero-temperature limit is $K_x^-(0)\Nf=-1/(\pi\kF)$.
In the high-temperature limit, the ideal occupations give
$f_x\sim-2\pi n(1+\zeta^2)/T$, and hence
$K_x^-(0)\Nf\sim-2/(\pi\thetared\kF)$.
At $\thetared=8$, the exchange reference differs by about $2.3\%$ from this
leading term.
Table~\ref{tab:exchange-reference} compares it with the VDMC uniform
kernel at all four densities.
The positive correlation contribution reduces the exchange magnitude
by $21$--$48\%$ across these densities.
This cancellation ratio applies equally to the free-energy curvatures,
since both kernel contributions are multiplied by $n$.
Subtracting the exchange reference from the corrKSDT kernel in
Fig.~\ref{fig:chi_ratio} gives a negative correlation contribution at $\thetared=8$.
This contrasts with the positive values obtained from VDMC and GDB
at that temperature.

\begin{table}[!htbp]
    \caption{\label{tab:exchange-reference}
    Uniform exchange and XC kernels at $\thetared=8$.
    The last column gives the percentage of the exchange magnitude
    canceled by correlation, $K_c^-=K_{xc}^{-,\mathrm{VDMC}}-K_x^-$.
    Parentheses denote propagated combined uncertainties in the last digits.}
    \begin{ruledtabular}
        \begin{tabular}{crrr}
            $\rs$ & $K_x^-(0)\Nf$ & $K_{xc}^{-,\mathrm{VDMC}}(0)\Nf$ & $K_c^-/|K_x^-|$ (\%) \\
            \hline
            $1$   & $-0.0405083$  & $-0.03189(18)$                   & $21.28(43)$          \\
            $2$   & $-0.0810165$  & $-0.05404(49)$                   & $33.30(61)$          \\
            $3$   & $-0.121525$   & $-0.07073(41)$                   & $41.80(34)$          \\
            $4$   & $-0.162033$   & $-0.0836(10)$                    & $48.40(63)$          \\
        \end{tabular}
    \end{ruledtabular}
\end{table}

\begin{figure*}[!t]
    \centering
    \includegraphics[width=0.9\textwidth]{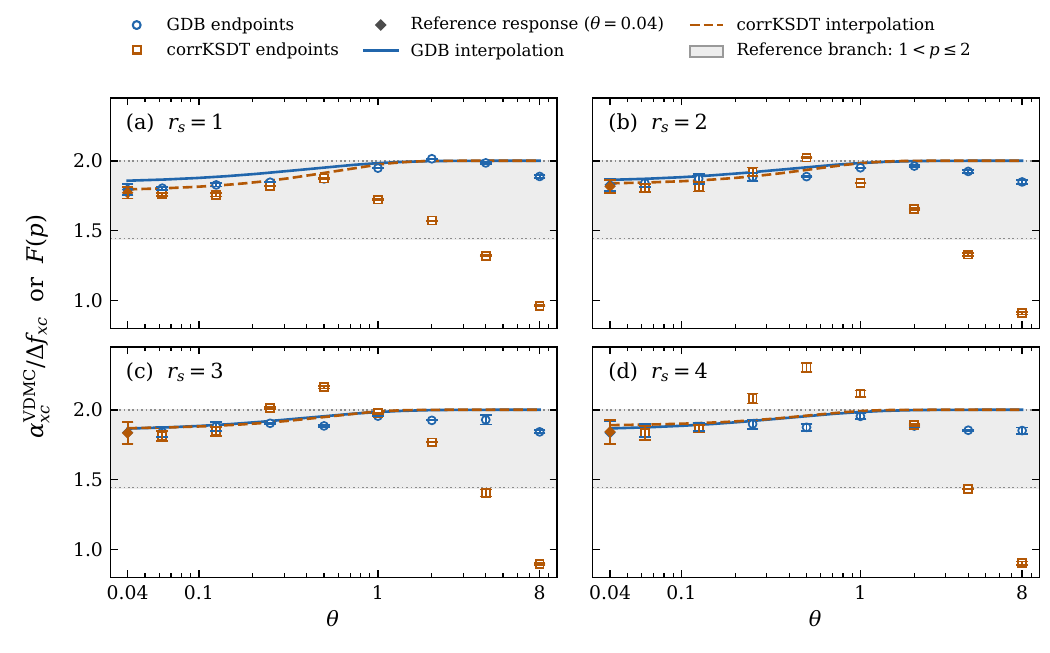}
    \caption{\label{fig:pdw-diagnostic}
    Response constraint on the PDW spin interpolation at fixed endpoint
    free energies, for $\rs=1,2,3,4$.
    Open circles and squares show $\alpha_{xc}^{\mathrm{VDMC}}/\Delta f_{xc}$
    using GDB and corrKSDT endpoints, respectively; filled diamonds use
    the earlier response data at $\thetared=0.04$ from
    Ref.~\cite{ChenHaule2019}, with the same endpoint color convention.
    Solid and dashed curves give $F(p)$ from GDB and corrKSDT, respectively.
    Shading marks $1/\ln2<F(p)\leq2$, the range corresponding to $1<p\leq2$.
    Error bars propagate the present combined response uncertainties
    (Sec.~\ref{sec:methods}) or the statistical uncertainties of
    Ref.~\cite{ChenHaule2019}, with endpoint free energies held fixed.}
\end{figure*}

\subsection{Spin interpolation}

To test whether changes to the spin interpolation can reproduce the VDMC
curvature, we hold the unpolarized and fully polarized endpoint free
energies fixed.
We evaluate both parametrizations with the PDW interpolation
~\cite{PerrotDharmawardana2000},
\begin{equation}
    \begin{aligned}
        f_{xc}(\rs,\thetared,\zeta)={} & f_{xc}^{0}(\rs,\thetared)                                      \\
                                       & +[f_{xc}^{1}(\rs,2^{-2/3}\thetared)-f_{xc}^{0}(\rs,\thetared)]
        \\[-2pt]
                                       & \qquad\times\frac{(1+\zeta)^p+(1-\zeta)^p-2}{2^p-2},
    \end{aligned}
    \label{eq:pdw}
\end{equation}
where $p=p(\rs,\thetared)$ is the spin-interpolation exponent, and
$f_{xc}^{0}$ and $f_{xc}^{1}$ are the unpolarized and fully polarized
endpoint parametrizations.
Each endpoint uses the Fermi temperature for its own polarization.
The fully polarized Fermi temperature is $2^{2/3}$ times the unpolarized
value, so the argument $2^{-2/3}\thetared$ keeps the physical temperature
fixed at the same density.

For GDB we use the endpoint functions and interpolation parameters of
Ref.~\cite{Groth2017}. For corrKSDT we combine the corrected unpolarized
endpoint of Ref.~\cite{Karasiev2018} with the fully polarized endpoint
and interpolation exponent of Ref.~\cite{KSDT2014}.
At fixed density and temperature, only the interpolation factor depends
on $\zeta$. Its second derivative at zero polarization is
$F(p)=2p(p-1)/(2^p-2)$, giving $\alpha_{xc}=\Delta f_{xc}F(p)$,
where $\Delta f_{xc}=f_{xc}^{1}(\rs,2^{-2/3}\thetared)-f_{xc}^{0}(\rs,\thetared)$.
The analytic curvature was checked against a symmetric finite difference
in the polarization.

Using $\alpha_{xc}^{\mathrm{VDMC}}=nK_{xc}^{-,\mathrm{VDMC}}(0)$ from
the uniform-response data in Fig.~\ref{fig:chi_ratio}, we form the ratio
$\alpha_{xc}^{\mathrm{VDMC}}/\Delta f_{xc}$ for each endpoint choice.
This ratio is the interpolation curvature factor required to reproduce
the calculated response with those endpoints fixed.

Figure~\ref{fig:pdw-diagnostic} compares these response-derived factors
with those of the original parametrizations.
At $\thetared=8$, both original parametrizations have $p\simeq2$ and
$F(p)\simeq2$, while the factors required by the VDMC response differ
substantially between the two endpoint choices.
Since the response numerator is shared, this separation reflects
the different polarized--unpolarized free-energy differences.
These endpoint differences therefore account for most of the
high-temperature spin-curvature discrepancy between GDB and corrKSDT.
Replacing only the corrected unpolarized endpoint with the original
KSDT form~\cite{KSDT2014} reduces the curvature magnitude at $\thetared=4,8$.
At $\thetared=8$, the original KSDT values nevertheless remain larger
in magnitude than GDB at all four densities.

For a conditional test of the interpolation, we consider the monotonic
branch $1<p\leq2$.
This interval contains the exponents used by both reference parametrizations
over the studied range, including their high-temperature limit $p=2$.
On this branch, $1/\ln2<F(p)\leq2$, so a response-derived ratio within
this interval determines a unique matching exponent.
At $\thetared=4,8$, the central values admit such an exponent with GDB
endpoints but lie below the interval with corrKSDT endpoints at all four densities.
At lower temperatures, the GDB-endpoint central values remain within
the interval except at $\rs=1,\thetared=2$, where the ratio lies slightly
above two; several corrKSDT-endpoint values exceed the upper boundary.

A consistent thermal spin parametrization must reproduce both the
response curvature and free energies at finite polarization, together
with the appropriate temperature limits.

\bibliography{references}

@article{KSDT2014,
  author  = {Karasiev, V. V. and Sjostrom, T. and Dufty, J. and Trickey, S. B.},
  title   = {Accurate Homogeneous Electron Gas Exchange-Correlation Free Energy for Local Spin-Density Calculations},
  journal = {Phys. Rev. Lett.},
  volume  = {112},
  pages   = {076403},
  year    = {2014},
  doi     = {10.1103/PhysRevLett.112.076403}
}

@article{Groth2017,
  author  = {Groth, S. and Dornheim, T. and Sjostrom, T. and Malone, F. D. and Foulkes, W. M. C. and Bonitz, M.},
  title   = {{\it Ab initio} Exchange-Correlation Free Energy of the Uniform Electron Gas at Warm Dense Matter Conditions},
  journal = {Phys. Rev. Lett.},
  volume  = {119},
  pages   = {135001},
  year    = {2017},
  doi     = {10.1103/PhysRevLett.119.135001}
}

@article{Karasiev2018,
  author  = {Karasiev, V. V. and Dufty, J. W. and Trickey, S. B.},
  title   = {Nonempirical Semilocal Free-Energy Density Functional for Matter under Extreme Conditions},
  journal = {Phys. Rev. Lett.},
  volume  = {120},
  pages   = {076401},
  year    = {2018},
  doi     = {10.1103/PhysRevLett.120.076401}
}

@article{Karasiev2019,
  author  = {Karasiev, V. V. and Trickey, S. B. and Dufty, J. W.},
  title   = {Status of Free-Energy Representations for the Homogeneous Electron Gas},
  journal = {Phys. Rev. B},
  volume  = {99},
  pages   = {195134},
  year    = {2019},
  doi     = {10.1103/PhysRevB.99.195134}
}

@article{PerrotDharmawardana2000,
  author  = {Perrot, F. and Dharma-wardana, M. W. C.},
  title   = {Spin-Polarized Electron Liquid at Arbitrary Temperatures: Exchange-Correlation Energies, Electron-Distribution Functions, and the Static Response Functions},
  journal = {Phys. Rev. B},
  volume  = {62},
  pages   = {16536},
  year    = {2000},
  doi     = {10.1103/PhysRevB.62.16536}
}

@article{PerdewWang1992,
  author  = {Perdew, J. P. and Wang, Y.},
  title   = {Accurate and Simple Analytic Representation of the Electron-Gas Correlation Energy},
  journal = {Phys. Rev. B},
  volume  = {45},
  pages   = {13244},
  year    = {1992},
  doi     = {10.1103/PhysRevB.45.13244}
}

@article{Spink2013,
  author  = {Spink, G. G. and Needs, R. J. and Drummond, N. D.},
  title   = {Quantum Monte Carlo Study of the Three-Dimensional Spin-Polarized Homogeneous Electron Gas},
  journal = {Phys. Rev. B},
  volume  = {88},
  pages   = {085121},
  year    = {2013},
  doi     = {10.1103/PhysRevB.88.085121}
}

@article{Brown2013,
  author  = {Brown, E. W. and Clark, B. K. and DuBois, J. L. and Ceperley, D. M.},
  title   = {Path-Integral Monte Carlo Simulation of the Warm Dense Homogeneous Electron Gas},
  journal = {Phys. Rev. Lett.},
  volume  = {110},
  pages   = {146405},
  year    = {2013},
  doi     = {10.1103/PhysRevLett.110.146405}
}

@article{Groth2017response,
  author  = {Groth, S. and Dornheim, T. and Bonitz, M.},
  title   = {Configuration path integral Monte Carlo approach to the static density response of the warm dense electron gas},
  journal = {J. Chem. Phys.},
  volume  = {147},
  pages   = {164108},
  year    = {2017},
  doi     = {10.1063/1.4999907}
}

@article{Schoof2015,
  author  = {Schoof, T. and Groth, S. and Vorberger, J. and Bonitz, M.},
  title   = {Ab Initio Thermodynamic Results for the Degenerate Electron Gas at Finite Temperature},
  journal = {Phys. Rev. Lett.},
  volume  = {115},
  pages   = {130402},
  year    = {2015},
  doi     = {10.1103/PhysRevLett.115.130402}
}

@article{Dornheim2016TDL,
  author  = {Dornheim, T. and Groth, S. and Sjostrom, T. and Malone, F. D. and Foulkes, W. M. C. and Bonitz, M.},
  title   = {Ab Initio Quantum Monte Carlo Simulation of the Warm Dense Electron Gas in the Thermodynamic Limit},
  journal = {Phys. Rev. Lett.},
  volume  = {117},
  pages   = {156403},
  year    = {2016},
  doi     = {10.1103/PhysRevLett.117.156403}
}

@article{Malone2016,
  author  = {Malone, F. D. and Blunt, N. S. and Brown, E. W. and Lee, D. K. K. and Spencer, J. S. and Foulkes, W. M. C. and Shepherd, J. J.},
  title   = {Accurate Exchange-Correlation Energies for the Warm Dense Electron Gas},
  journal = {Phys. Rev. Lett.},
  volume  = {117},
  pages   = {115701},
  year    = {2016},
  doi     = {10.1103/PhysRevLett.117.115701}
}

@article{Dornheim2017PBresponse,
  author  = {Dornheim, T. and Groth, S. and Vorberger, J. and Bonitz, M.},
  title   = {Permutation-blocking path-integral Monte Carlo approach to the static density response of the warm dense electron gas},
  journal = {Phys. Rev. E},
  volume  = {96},
  pages   = {023203},
  year    = {2017},
  doi     = {10.1103/PhysRevE.96.023203}
}

@article{Dornheim2019LFC,
  author  = {Dornheim, T. and Vorberger, J. and Groth, S. and Hoffmann, N. and Moldabekov, Zh. A. and Bonitz, M.},
  title   = {The static local field correction of the warm dense electron gas: An {\it ab initio} path integral Monte Carlo study and machine learning representation},
  journal = {J. Chem. Phys.},
  volume  = {151},
  pages   = {194104},
  year    = {2019},
  doi     = {10.1063/1.5123013}
}

@article{Dornheim2018,
  author  = {Dornheim, T. and Groth, S. and Bonitz, M.},
  title   = {The Uniform Electron Gas at Warm Dense Matter Conditions},
  journal = {Phys. Rep.},
  volume  = {744},
  pages   = {1},
  year    = {2018},
  doi     = {10.1016/j.physrep.2018.04.001}
}

@article{Dornheim2022spin,
  author  = {Dornheim, T. and Vorberger, J. and Moldabekov, Z. A. and Tolias, P.},
  title   = {Spin-Resolved Density Response of the Warm Dense Electron Gas},
  journal = {Phys. Rev. Research},
  volume  = {4},
  pages   = {033018},
  year    = {2022},
  doi     = {10.1103/PhysRevResearch.4.033018}
}

@misc{Dornheim2022rodare,
  author       = {Dornheim, Tobias},
  title        = {Data publication: Spin-resolved density response of the warm dense electron gas},
  year         = {2022},
  howpublished = {RODARE dataset},
  doi          = {10.14278/rodare.1698},
  url          = {https://doi.org/10.14278/rodare.1698}
}

@misc{UEGSpinData,
  author       = {Hou, Pengcheng and Li, Zhiyi and Deng, Youjin and Chen, Kun},
  title        = {Static Spin Exchange-Correlation Kernel of the Uniform Electron Gas at Finite Temperature: Data Release},
  year         = {2026},
  howpublished = {GitHub repository},
  note         = {Data version 9f8dfce},
  url          = {https://github.com/houpc/UEG_spinXC/tree/9f8dfce2d05c9230c508091b86582fba0760d51d}
}

@article{Hou2022,
  author  = {Hou, P. and Wang, B.-Z. and Haule, K. and Deng, Y. and Chen, K.},
  title   = {Exchange-Correlation Effect in the Charge Response of a Warm Dense Electron Gas},
  journal = {Phys. Rev. B},
  volume  = {106},
  pages   = {L081126},
  year    = {2022},
  doi     = {10.1103/PhysRevB.106.L081126}
}

@article{Li2025,
  author  = {Li, Z. and Hou, P. and Deng, Y. and Chen, K.},
  title   = {Matsubara-frequency-resolved spin exchange-correlation kernel for the three-dimensional uniform electron gas},
  journal = {Phys. Rev. B},
  volume  = {111},
  pages   = {155132},
  year    = {2025},
  doi     = {10.1103/PhysRevB.111.155132}
}

@article{ChenHaule2019,
  author  = {Chen, K. and Haule, K.},
  title   = {A Combined Variational and Diagrammatic Quantum Monte Carlo Approach to the Many-Electron Problem},
  journal = {Nat. Commun.},
  volume  = {10},
  pages   = {3725},
  year    = {2019},
  doi     = {10.1038/s41467-019-11708-6}
}

@article{KukkonenOverhauser1979,
  author  = {Kukkonen, C. A. and Overhauser, A. W.},
  title   = {Electron-Electron Interaction in Simple Metals},
  journal = {Phys. Rev. B},
  volume  = {20},
  pages   = {550},
  year    = {1979},
  doi     = {10.1103/PhysRevB.20.550}
}

@article{KukkonenChen2021,
  author  = {Kukkonen, C. A. and Chen, K.},
  title   = {Quantitative Electron-Electron Interaction Using Local-Field Factors from Quantum Monte Carlo Calculations},
  journal = {Phys. Rev. B},
  volume  = {104},
  pages   = {195142},
  year    = {2021},
  doi     = {10.1103/PhysRevB.104.195142}
}

@article{RichardsonAshcroft1994,
  author  = {Richardson, C. F. and Ashcroft, N. W.},
  title   = {Dynamical Local-Field Factors and Effective Interactions in
             the Three-Dimensional Electron Liquid},
  journal = {Phys. Rev. B},
  volume  = {50},
  pages   = {8170},
  year    = {1994},
  doi     = {10.1103/PhysRevB.50.8170}
}

@article{KaplanKukkonen2023,
  author  = {Kaplan, A. D. and Kukkonen, C. A.},
  title   = {QMC-Consistent Static Spin and Density Local Field Factors for the Uniform Electron Gas},
  journal = {Phys. Rev. B},
  volume  = {107},
  pages   = {L201120},
  year    = {2023},
  doi     = {10.1103/PhysRevB.107.L201120}
}

@article{Mermin1965,
  author  = {Mermin, N. D.},
  title   = {Thermal Properties of the Inhomogeneous Electron Gas},
  journal = {Phys. Rev.},
  volume  = {137},
  pages   = {A1441},
  year    = {1965},
  doi     = {10.1103/PhysRev.137.A1441}
}

@article{Stoitsov1988,
  author  = {Stoitsov, M. V. and Petkov, I. Zh.},
  title   = {Density Functional Theory at Finite Temperatures},
  journal = {Ann. Phys. (N.Y.)},
  volume  = {184},
  pages   = {121},
  year    = {1988},
  doi     = {10.1016/0003-4916(88)90271-0}
}

@article{PribramJones2016,
  author  = {Pribram-Jones, Aurora and Grabowski, Paul E. and Burke, Kieron},
  title   = {Thermal Density Functional Theory: Time-Dependent Linear Response and Approximate Functionals from the Fluctuation-Dissipation Theorem},
  journal = {Phys. Rev. Lett.},
  volume  = {116},
  pages   = {233001},
  year    = {2016},
  doi     = {10.1103/PhysRevLett.116.233001}
}

@article{RungeGross1984,
  author  = {Runge, E. and Gross, E. K. U.},
  title   = {Density-Functional Theory for Time-Dependent Systems},
  journal = {Phys. Rev. Lett.},
  volume  = {52},
  pages   = {997},
  year    = {1984},
  doi     = {10.1103/PhysRevLett.52.997}
}

@article{GrossKohn1985,
  author  = {Gross, E. K. U. and Kohn, W.},
  title   = {Local Density-Functional Theory of Frequency-Dependent Linear Response},
  journal = {Phys. Rev. Lett.},
  volume  = {55},
  pages   = {2850},
  year    = {1985},
  doi     = {10.1103/PhysRevLett.55.2850}
}

@article{Bonitz2020,
  author  = {Bonitz, M. and Dornheim, T. and Moldabekov, Zh. A. and Zhang, S. and Hamann, P. and Kahlert, H. and Filinov, A. and Ramakrishna, K. and Vorberger, J.},
  title   = {Ab Initio Simulation of Warm Dense Matter},
  journal = {Phys. Plasmas},
  volume  = {27},
  pages   = {042710},
  year    = {2020},
  doi     = {10.1063/1.5143225}
}

@article{Moldabekov2023,
  author  = {Moldabekov, Zh. A. and B{\"o}hme, M. and Vorberger, J. and Blaschke, D. and Dornheim, T.},
  title   = {Ab Initio Static Exchange--Correlation Kernel across Jacob's Ladder without Functional Derivatives},
  journal = {J. Chem. Theory Comput.},
  volume  = {19},
  pages   = {1286--1299},
  year    = {2023},
  doi     = {10.1021/acs.jctc.2c01180}
}

@article{Hurricane2016,
  author  = {Hurricane, O. A. and Callahan, D. A. and Casey, D. T. and others},
  title   = {Inertially Confined Fusion Plasmas Dominated by Alpha-Particle Self-Heating},
  journal = {Nat. Phys.},
  volume  = {12},
  pages   = {800},
  year    = {2016},
  doi     = {10.1038/nphys3720}
}

@article{SaumonGuillot2004,
  author  = {Saumon, D. and Guillot, T.},
  title   = {Shock Compression of Deuterium and the Interiors of Jupiter and Saturn},
  journal = {Astrophys. J.},
  volume  = {609},
  pages   = {1170},
  year    = {2004},
  doi     = {10.1086/421257}
}

@article{Schottler2018,
  author  = {Sch{\"o}ttler, M. and Redmer, R.},
  title   = {Ab Initio Calculation of the Miscibility Diagram for Hydrogen--Helium Mixtures},
  journal = {Phys. Rev. Lett.},
  volume  = {120},
  pages   = {115703},
  year    = {2018},
  doi     = {10.1103/PhysRevLett.120.115703}
}

@article{Kraus2017,
  author  = {Kraus, D. and Vorberger, J. and Pak, A. and others},
  title   = {Formation of Diamonds in Laser-Compressed Hydrocarbons at Planetary Interior Conditions},
  journal = {Nat. Astron.},
  volume  = {1},
  pages   = {606},
  year    = {2017},
  doi     = {10.1038/s41550-017-0219-9}
}

@article{Stoner1938,
  author  = {Stoner, E. C.},
  title   = {Collective Electron Ferromagnetism},
  journal = {Proc. R. Soc. A},
  volume  = {165},
  pages   = {372},
  year    = {1938},
  doi     = {10.1098/rspa.1938.0066}
}

@article{Wolff1960,
  author  = {Wolff, P. A.},
  title   = {Spin Susceptibility of an Electron Gas},
  journal = {Phys. Rev.},
  volume  = {120},
  pages   = {814},
  year    = {1960},
  doi     = {10.1103/PhysRev.120.814}
}

@book{GiulianiVignale2005,
  author    = {Giuliani, G. and Vignale, G.},
  title     = {Quantum Theory of the Electron Liquid},
  publisher = {Cambridge University Press},
  year      = {2005},
  address   = {Cambridge}
}

@article{Dornheim2021SpinPolarizedMomentum,
  author  = {Dornheim, Tobias and Vorberger, Jan and Militzer, Burkhard and Moldabekov, Zhandos A.},
  title   = {Momentum Distribution of the Uniform Electron Gas at Finite Temperature: Effects of Spin Polarization},
  journal = {Phys. Rev. E},
  volume  = {104},
  pages   = {055206},
  year    = {2021},
  doi     = {10.1103/PhysRevE.104.055206}
}

@article{Moldabekov2022SpinBenchmark,
  author  = {Moldabekov, Zhandos and Dornheim, Tobias and Vorberger, Jan and Cangi, Attila},
  title   = {Benchmarking Exchange-Correlation Functionals in the Spin-Polarized Inhomogeneous Electron Gas under Warm Dense Conditions},
  journal = {Phys. Rev. B},
  volume  = {105},
  pages   = {035134},
  year    = {2022},
  doi     = {10.1103/PhysRevB.105.035134}
}

@article{AroraMoudgil2025,
  author  = {Arora, P. and Moudgil, R. K.},
  title   = {A Dynamic Mean-Field Study of Spin-Resolved Static Density Response of Warm-Dense Homogeneous Electron Gas},
  journal = {Eur. Phys. J. B},
  volume  = {98},
  pages   = {68},
  year    = {2025},
  doi     = {10.1140/epjb/s10051-025-00916-3}
}

\end{document}